\documentclass[conf]{new-aiaa}
\usepackage[utf8]{inputenc}

\usepackage{subfigmat}
\usepackage{color,bm}

\usepackage{graphicx}
\usepackage{amsmath}
\usepackage{longtable,tabularx}
\title{High-Fidelity Semianalytical Theory for a Low Lunar Orbit}

\author{Juan F\'elix San-Juan\footnote{Professor, Scientific Computing Group (GRUCACI), juanfelix.sanjuan@unirioja.es, Senior Member AIAA.},
Rosario L\'opez\footnote{Assistant Professor, Scientific Computing Group (GRUCACI), rosario.lopez@unirioja.es.}
and Iv\'an P\'erez\footnote{Associate Professor, Scientific Computing Group (GRUCACI), ivan.perez@unirioja.es.}}
\affil{University of La Rioja, Logro\~{n}o, La Rioja, 26006, Spain}

\begin{document}

\maketitle

\section{Introduction}
\lettrine{S}{ince} the 1990s,  there has been a renewed interest in the  exploration of the Moon. A significant number of lunar missions   have been and are being  conducted by the  United States of America, Russia, Europe, Japan, India, and  China. Most of these missions require  low-altitude orbits. For this reason, looking for  ideal orbits that minimize or cancel the undesirable effects of some perturbations which can modify     spacecraft  orbits continue to be a current problem. Classically, these ideal orbits have been  known as frozen orbits.  A  historical review of the frozen-orbit concept can be found in  \cite{cof1994_frozearthlike} and the references cited therein. 

Basically, from the mathematical point of view, a frozen orbit corresponds to an  equilibrium of a system of differential equations in which the influence of the short- and medium-period terms has been removed. 
The reduced or double-averaged  system only maintains the long-period dynamics. This  system of differential equations  can be derived in two ways. First, the problem can be formulated using the well-known Lagrange planetary equations, where the short- and medium-period terms can be removed by classical averaging techniques \cite{coo1966_pertnearcirc,ros1992_highzfroz,kie1992_frozj2j3,coo1992_nearcirc, fol1998_lunarprospfroz,par1994_lunarmap,fol2006_lunarfroz}. Second, the problem can also be expressed in Hamiltonian form, in which case the elimination process can be carried out by  sophisticated methods based on Lie transforms \cite{cof1994_frozearthlike,koz1963_lunar,lar2009gru_lunar50z3body,lar2011gru_longlunar}, which allow determining easily the transformations between mean and osculating elements.

In general, these ideas have been  employed extensively for  mission-design studies. When they are applied to the case of an orbiter around the Earth,  only a few zonal harmonic coefficients of the Earth's gravitational potential are enough to allow  characterizing the phase-space structure of the full problem \cite{cof1994_frozearthlike}. However, in the case of  real mission-design analyses of low lunar orbiters, it is necessary to consider a full perturbation model which must include  both a high-resolution lunar gravitational-field model  and the  third-body attraction caused by the Earth \cite{rus2007gru_repgtlunar}. As a reference, Roncoli \cite{ron2005_lunarconstants} recommends a $50\times 50$ lunar gravitational-field model as the minimum for orbits with altitudes below 100 km, resolution later confirmed by Lara et al. in \cite{lar2009gru_lunar50z3body,lar2011gru_longlunar}.

In this context, we consider it necessary to revisit and extend the semi-analytical models developed in \cite{san2008gru_lunar_aas,san2010gru_lunarrev_aas}, whose aim was to test the feasibility of perturbation theories based on Lie transforms, and therefore constituted mere academic proof-of-concept studies that served the author to develop specific software for this task\footnote{Using this software, results in \cite{san2008gru_lunar_aas} where later validated by the coworkers of the first author of this paper \cite{aba2009_frozlunar}.}. Indeed, since the first high-resolution lunar gravitational fields were released \cite{kon1994_lunargravit}, it  is well known that high-inclination, low-altitude lunar orbits behave quite differently from what lower-degree models predict.

In the present work, both the symbolic-manipulation software developed by the authors of this article and our expertise from previous models are applied to the analysis  of an orbiter in a low-altitude near-circular orbit around the Moon, perturbed by the combined effect of a  $50\times 50$ lunar gravitational-field model and the  third-body attraction caused by the Earth. Our objective is to develop a semi-analytical theory that can be integrated both in a symbolic software tool for the computation of real low-altitude frozen orbits and in a semi-analytical propagator for low-altitude orbits. By means of two Lie transforms, we remove the short- and medium-period terms from the original problem.  As a result, the double-averaged problem  only depends on the argument of the periapsis,   and then its long-term behavior can be studied.  Finally, we analyze the influence of the zonal coefficients and  the third-body attraction from the Earth in the phase-space structure of the double-averaged problem.  The resulting  analytical expressions can be used to locate frozen orbits in the double-averaged problem, as well as to  perform all the required computations in order to transform the frozen conditions into  a near-periodic orbit for the original problem.

\section{Dynamical model}

The following Hamiltonian  describes the motion of an orbiter around the Moon, perturbed  by its gravitational force, $\mathcal{V}_{M}$, and by the third-body attraction due to the Earth $\mathcal{V}_{3b}$, under the Hill hypothesis, that is, assuming that the Moon is in circular orbit around the Earth and that the Earth is moving in a circular orbit in synchronization with the lunar rotation:

\begin{equation}\label{ham1}
\mathcal{H} = \frac{1}{2}\bm{X} \cdot \bm{X} - \bm{\nu} \cdot (\bm{x} \times \bm{X})
+ \mathcal{V}_{M} + \mathcal{V}_{3b}
\end{equation}

\noindent
In this equation, $\bm{x}$ are coordinates and $\bm{X}$ their conjugate momenta, and $\bm{\nu} $ represents the angular velocity vector of the Moon. The lunar gravitational potential, for which only the order and degree $50$   is taken into account,  is given in selenocentric spherical coordinates by

\begin{eqnarray}\label{poten}
\mathcal{V}_{M} = - \frac{\mu}{r} \sum_{n = 0}^{50} \left(\frac{\alpha}{r} \right)^{n}
   \sum_{0\le m \le n} ( C_{nm} \cos m\lambda + S_{nm} \sin m\lambda)
   P_{n}^{m}(\sin \beta)
\end{eqnarray}

\noindent
where $\alpha$ is the equatorial radius of the Moon,  $\mu$ the gravitational parameter for the Moon, $C_{nm} $ and $S_{nm} $  the lunar harmonic  coefficients, and $P_{n}^{m}$ the Legendre function of degree $n$ and order $m$. For the third-body attraction, only the dominant  effect is considered, which is given by

\begin{equation}
 \mathcal{V}_{3b} = \frac{\nu^2}{2} (r^2 - 3 x^2)
 \end{equation}

For the  low-altitude orbit case,  approximately
below $200$ km, the Hamiltonian \eqref{ham1} can be rewritten as

\begin{equation}\label{ham2}
\mathcal{H} = \mathcal{H}_K + \epsilon \mathcal{H}_{\nu} + 
\frac{\epsilon^2}{2!} \left(\mathcal{V}_{M} + \mathcal{V}_{3b}\right)
\end{equation}

\noindent
where $\mathcal{H}_K$ and  $\mathcal{H}_{\nu}$  correspond to the Kepler problem and the Coriolis effect, respectively. $\epsilon$ represents a small parameter \cite{koz1963_lunar,gar1965_tesspert} which is defined as $\nu/n$, where $n$ is the mean motion of the orbiter.

In order to reveal the qualitative behavior of the dynamic system given in \eqref{ham2}, we  use two Lie transforms: the elimination of the parallax \cite{dep1981_parallax,san2013gru_depparallaxrev,lar2014gru_parallax_aas}, in order to reduce the number of terms of the transformed Hamiltonian, and the double normalization \cite{dep1982_delaunaynorm,san2015gru_arbitdeprit}, to remove simultaneously the short- and medium-period terms. As a consequence, the system is reduced to an integrable one governed by a truncated second-order closed-form Hamiltonian.
In Delaunay variables the transformed Hamiltonian yields

\begin{equation}\label{ham3}
\mathcal{H}''= \mathcal{H}''_0 + \epsilon \mathcal{H}''_1 + \frac{\epsilon^2}{2!} \mathcal{H}''_2
\end{equation}

\noindent
with $\mathcal{H}''_0 = -\mu^2/(2L''^2)$,  $\mathcal{H}''_1= - (\nu/\epsilon) H''$, and
$\mathcal{H}''_2 =   \mathcal{H}''_2 (\epsilon,J_n=-C_{n0},\mu,\alpha;g'',L'',G'',H'')$. It only depends on the long-period terms and the dynamic and physical parameters, which are handled as symbolic constants. The total number of terms of $\mathcal{H}''$ is $16\,103$, which reveals the current capabilities of both the hardware and the general-purpose symbolic-manipulation software, such as \textit{Mathematica}, as opposed to the usual situation not so many years ago, when specific software, for example the Poisson-series processors, was necessary in order to handle this type of expressions. From the generating function of the two Lie transforms, it is  easy to determine the transformations from mean to osculating and from osculating to mean elements. It is worth noting that the second order of this theory is equivalent to applying the classical averaging techniques twice over  Hamiltonian \eqref{ham1}.

The  semi-analytical theory has been developed using MathATESAT \cite{san2011gru_mathatesat}. It is a framework embedded in \textit{Mathematica} which allows applying perturbation methods based on Lie transforms and classical averaging techniques to perturbed Keplerian and oscillator systems. Thus, MathATESAT  benefits from  other \textit{Mathematica} capabilities  such as  visualization, resolution of systems of algebraic equations, numerical integration, and so forth. Another facility included in our system is the capability to export all the \textit{Mathematica} expressions involved in the analytical theory in order to evaluate them later in an independent  program.

\section{Phase-space structure: frozen orbits}

The phase-space structure of the double-reduced Hamiltonian \eqref{ham3} can be explored by using graphical, analytical, and numerical techniques \cite{cof1986_critincl_art,cof1990_paintphspace,coo1992_nearcirc,cof1994_frozearthlike,fer2007gru_boundsmainp}. This Hamiltonian is governed  by the following system of differential equations

\begin{eqnarray}\label{ham4}
\frac{dg''}{dt} & = & \frac{\partial \mathcal{H}''}{\partial G''}=\sum_{j=0}^{24} \mathcal{P}_{2j}^{g''} \cos 2j \,g''  + \sum_{j=0}^{23} \mathcal{Q}_{2j+1}^{g''} \sin (2j+1)g'' , \\[1ex]
\frac{dG''}{dt} & = &-\frac{\partial \mathcal{H}''}{\partial g''} = \sum_{j=0}^{23} \mathcal{P}_{2j+1}^{G''} \cos (2j+1)g''  +\sum_{j=1}^{24} \mathcal{Q}_{2j}^{G''} \sin 2j\,g'' \nonumber
\end{eqnarray}

\noindent
where $\mathcal{P}_{2j}^{g''}$, $\mathcal{Q}_{2j+1}^{g''}$,  $\mathcal{P}_{2j+1}^{G''} $ and $\mathcal{Q}_{2j}^{G''}$ depend on all the physical parameters, and on the momenta $L''$, $G''$ and $H''$. This system describes the averaged motion of the argument of the periapsis and the angular momentum. From Eqs. \eqref{ham4}, the global qualitative behavior can be described in function of the dynamics constraints $(L'',H'')$ obtained from Hamilton's equations.
Although the angle-action variables of Delaunay are best suited for the development of this semi-analytical theory, the more commonly used orbital elements seem more convenient so as to present some graphical results from which to draw conclusions. The conversion between both sets of variables can be easily done by means of the following expressions: $a=L^2/\mu$, \, $e=\sqrt{1-(G/L)^2}$, \, $i=\arccos (H/G)$, \, $\Omega=h$, \, $\omega=g$, \, $M=l$.

Since our goal is to analyze the conditions in which frozen orbits can exist, depending on both the number of zonal harmonics considered in the dynamical model and the perturbation introduced by the third-body attraction from the Earth, we will use the classical representation of families of frozen orbits in function of both the inclination and eccentricity. It is worth mentioning that generating the figures that will be presented next, which we have done with \textit{Mathematica}, requires a very intensive computational process, derived from the fact that the analytical expression that we need to evaluate comprises $159\,479$ terms. Moreover, in order not to lose accuracy in intermediate calculations, physical parameters have been rationalized so as to use rational arithmetic, and the working precision has been set to $100$ digits, thereby increasing further the computational burden.

In order to investigate the phase-space structure of the double-reduced Hamiltonian given in \eqref{ham3}, we determine the equilibrium solutions to  system \eqref{ham4}. These solutions correspond to the so-called \textit{frozen orbits}, which are characterized by keeping both the mean eccentricity and the argument of the periapsis constant. These orbits are determined by the conditions

\begin{equation} \label{eq_cond}
\frac{d g''}{d t}=\frac{d G''}{d t}=0
\end{equation}

It can be verified that $\cos g''=0$ constitutes a trivial solution to these equations, which leads to the well-known argument-of-the-periapsis values of $g=90^\circ$ and $g=270^\circ$ at which frozen orbits can be found \cite{coo1966_pertnearcirc}. Some other arguments of the periapsis have been obtained for low-order truncations of the gravitational potential in the case of the Earth \cite{cof1994_frozearthlike,lar2018gru_truncfroz_stardust}. Nevertheless, the huge magnitude of the expressions handled in this case would make it very complex to conduct an equivalent search.

Next, we present some graphics to illustrate the decisive effect that the high-order zonal coefficients of the lunar gravitational field exert on the dynamics of frozen orbits, which makes those harmonics indispensable in the process of searching for this type of orbits. Since the software we have developed for generating these plots is completely parameterized, it admits any gravitational models \cite{kon2001_gravitlunarprosp,mat2010_lunargravitselene,maz2010_150lunargravit,kon2013_lunargravit660}. We will use the LP150 model for this purpose.

\begin{figure}[!!htp]
\begin{center}
\includegraphics[scale=.7]{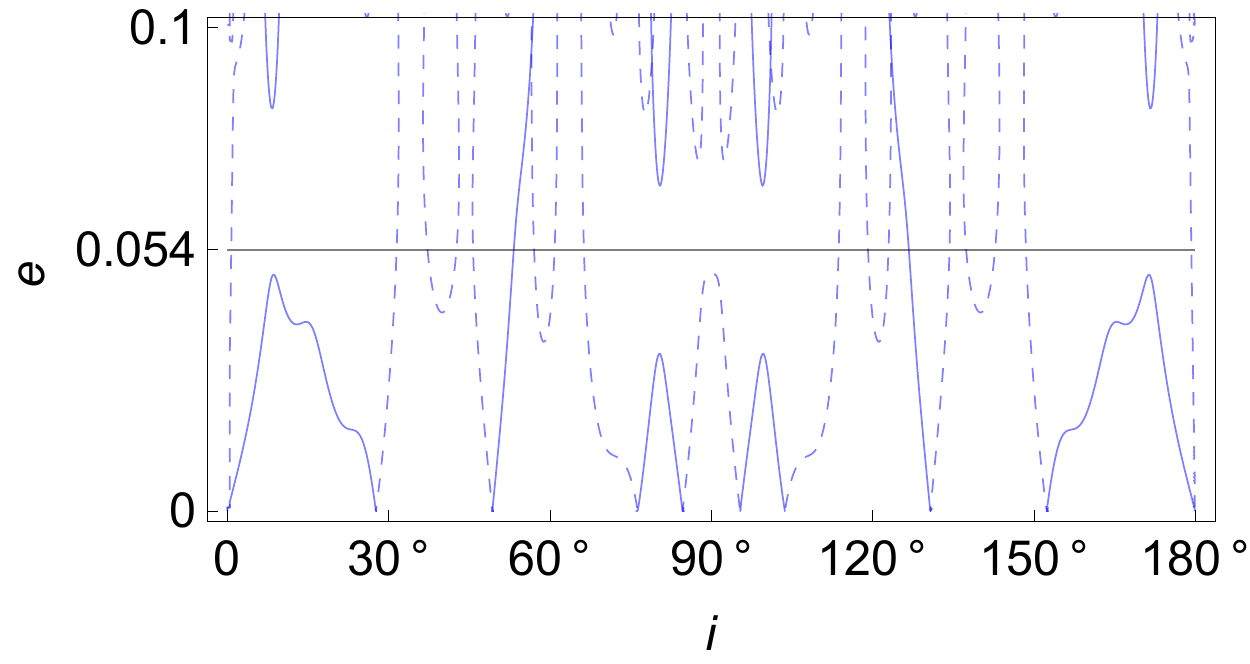}
\end{center} 
\caption{Characterization of frozen orbits for a mean semi-major axis of $\bm{1838}$ km.}
 \label{figure:fig1}
 \end{figure}

First, we characterize the complete set of families of frozen orbits that can be found at an altitude of $100$ km when a $50\times 50$ lunar gravitational-field model is considered, together with the third-body attraction from the Earth. With that aim, Fig.~\ref{figure:fig1} depicts the feasible couples of eccentricity and inclination values, which can correspond to an argument of the periapsis of either $90^\circ$, solid lines, or $270^\circ$, dashed lines. It is worth noting that there is an eccentricity value, $0.054$ for the mean semi-major axis of $1838$ km that we are considering, above which the periapsis altitude would be negative, therefore corresponding to unfeasible orbits that would impact on the lunar surface. Due to the fact that the averaged Hamiltonian only depends on the even powers of the cosine of the inclination, the plot is symmetrical about the inclination of $90^\circ$, as can be observed, which allows omitting inclinations from $90^\circ$ to $180^\circ$ from this kind of graphics for the sake of clarity, as we will do in some of the following figures. It should be noted that there are five possible values of inclination, $0^\circ$, $27.62^\circ$, $49.27^\circ$, $76.31^\circ$, and $84.64^\circ$, plus their corresponding symmetrical values for retrograde orbits, for which circular frozen orbits can be found. In all those cases, the circular orbit always implies a change in the argument of the periapsis between frozen orbits with lower and higher inclinations.

\begin{figure}[!!htp]
\begin{subfigmatrix}{3}
\subfigure[Argument of the periapsis $90^\circ$.]{\includegraphics{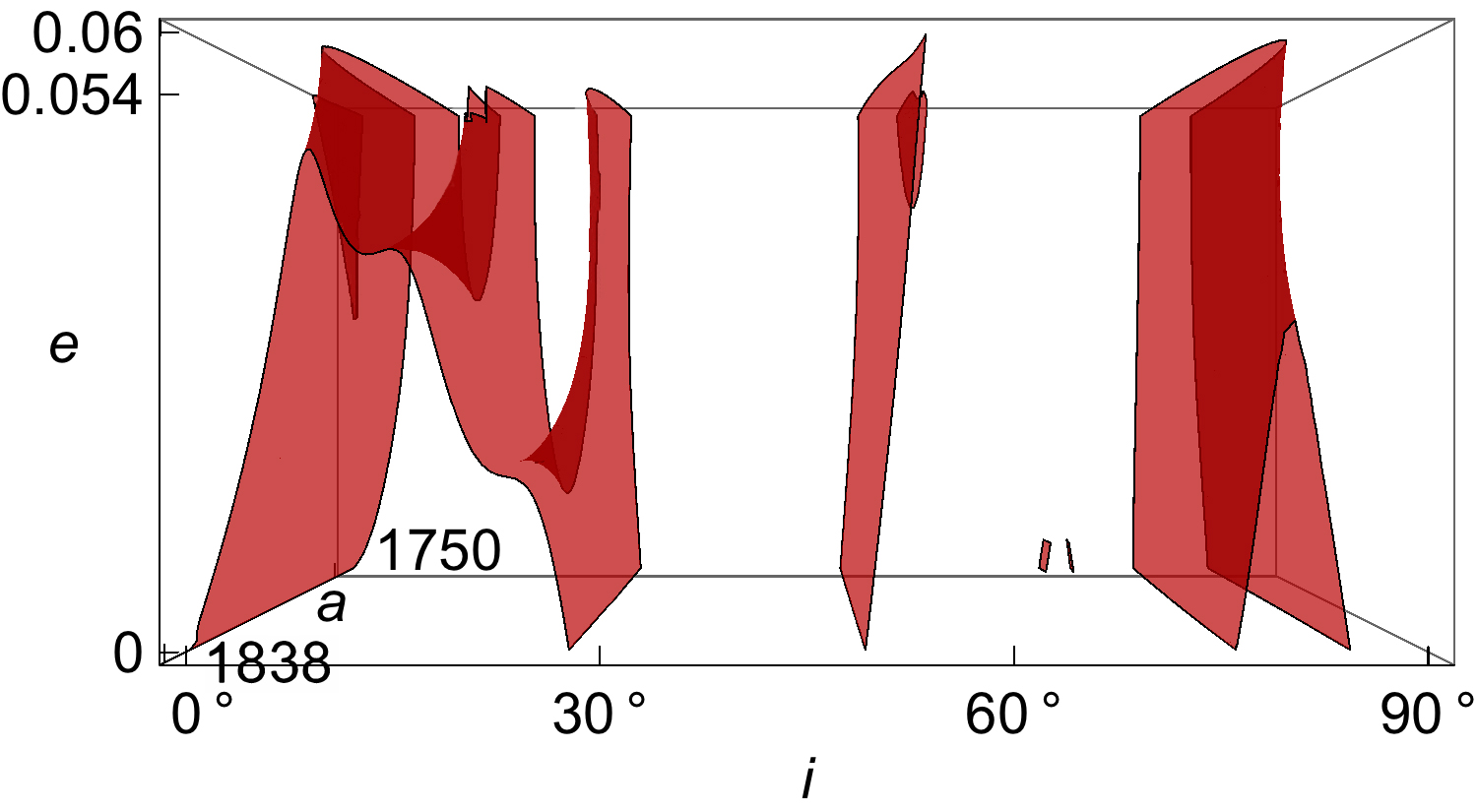}}
\subfigure[Argument of the periapsis $270^\circ$.]{\includegraphics{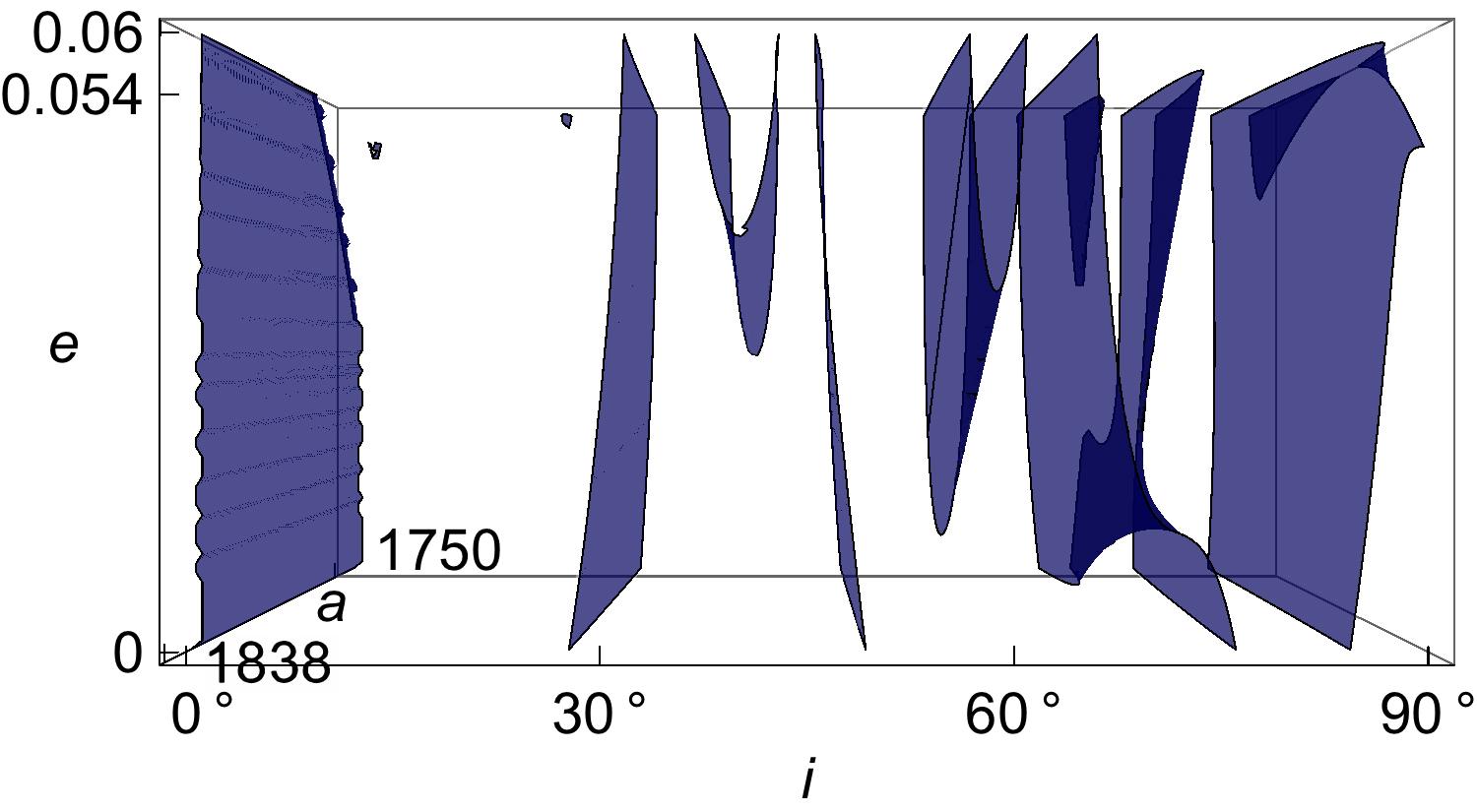}}
\subfigure[Both arguments of the periapsis, $90^\circ$ and $270^\circ$.]{\includegraphics{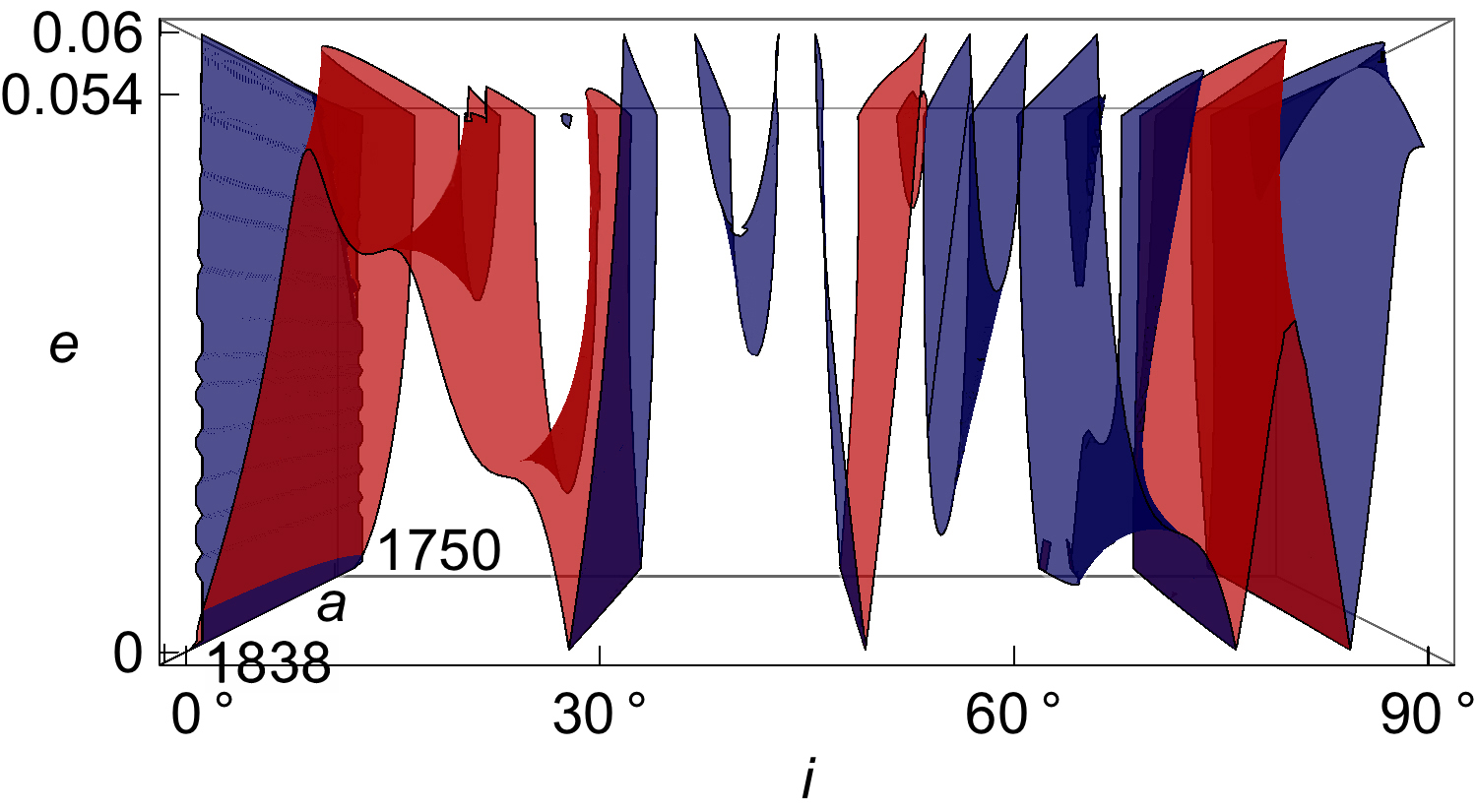}}
\end{subfigmatrix}
 \caption{Families of frozen orbits for semi-major axes between $\bm{1750}$ and $\bm{1838}$ km.
}
\label{figure:fig2}
 \end{figure}

The plot shown in Fig.~\ref{figure:fig1} can vary slightly for different values of semi-major axis within the range of low orbits. In order to illustrate that variation, Fig.~\ref{figure:fig2} extends Fig.~\ref{figure:fig1} by including a third axis that represents semi-major axis values from $1750$ to $1838$ km, thus corresponding to orbits with mean altitudes from $12$ to $100$ km. Red surfaces correspond to an argument of the periapsis of $90^\circ$ and blue surfaces to $270^\circ$. As previously mentioned, only inclinations from $0^\circ$ to $90^\circ$ have been plotted, since inclinations between $90^\circ$ and $180^\circ$ show the same patterns for retrograde orbits.

\begin{figure}[!!htp]
\begin{center}
\includegraphics[scale=.7]{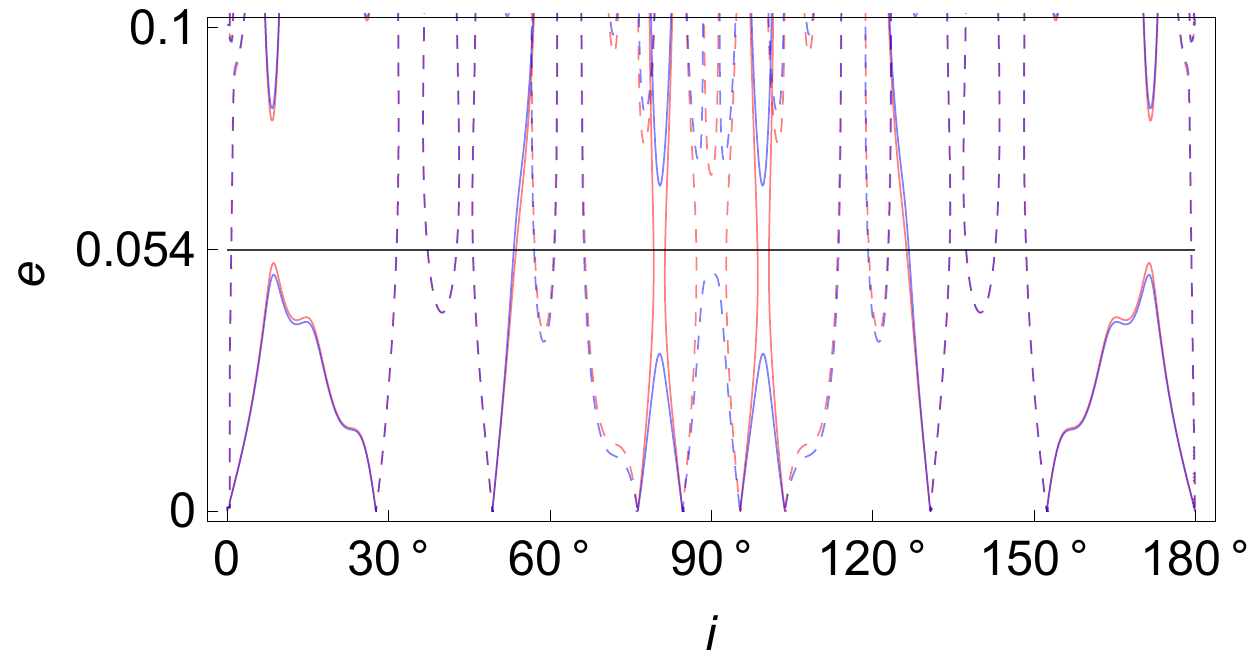}
\end{center} 
\caption{Families of frozen orbits for a mean semi-major axis of $\bm{1838}$ km.}
 \label{figure:fig3}
 \end{figure}

With the aim of analyzing separately the effects of both the lunar gravitational field and the third-body attraction from the Earth over the families of frozen orbits, Fig.~\ref{figure:fig3} includes a new plot with respect to Fig.~\ref{figure:fig1}, the red one, which represents the families of frozen orbits when the third-body effect is ignored. It is worth mentioning that this figure coincides with the central plot of Fig.~3 in Ref.~\cite{lar2011gru_longlunar}. As can be observed, the differences between the blue and the red plots are not significant for inclinations below approximately $70^\circ$ and their corresponding retrograde inclinations. Nevertheless, both plots exhibit important differences for frozen orbits near the polar region, for inclinations above  $70^\circ$ and the equivalent retrograde orbits. In these cases, considering the gravitational pull from the Earth becomes indispensable, since it can make the difference between a feasible frozen orbit and a non-viable orbit with an eccentricity over the impact limit.

\begin{figure}[!!htp]
\begin{subfigmatrix}{2}
\subfigure[\label{figure:fig4a} $J_2$.]{\includegraphics{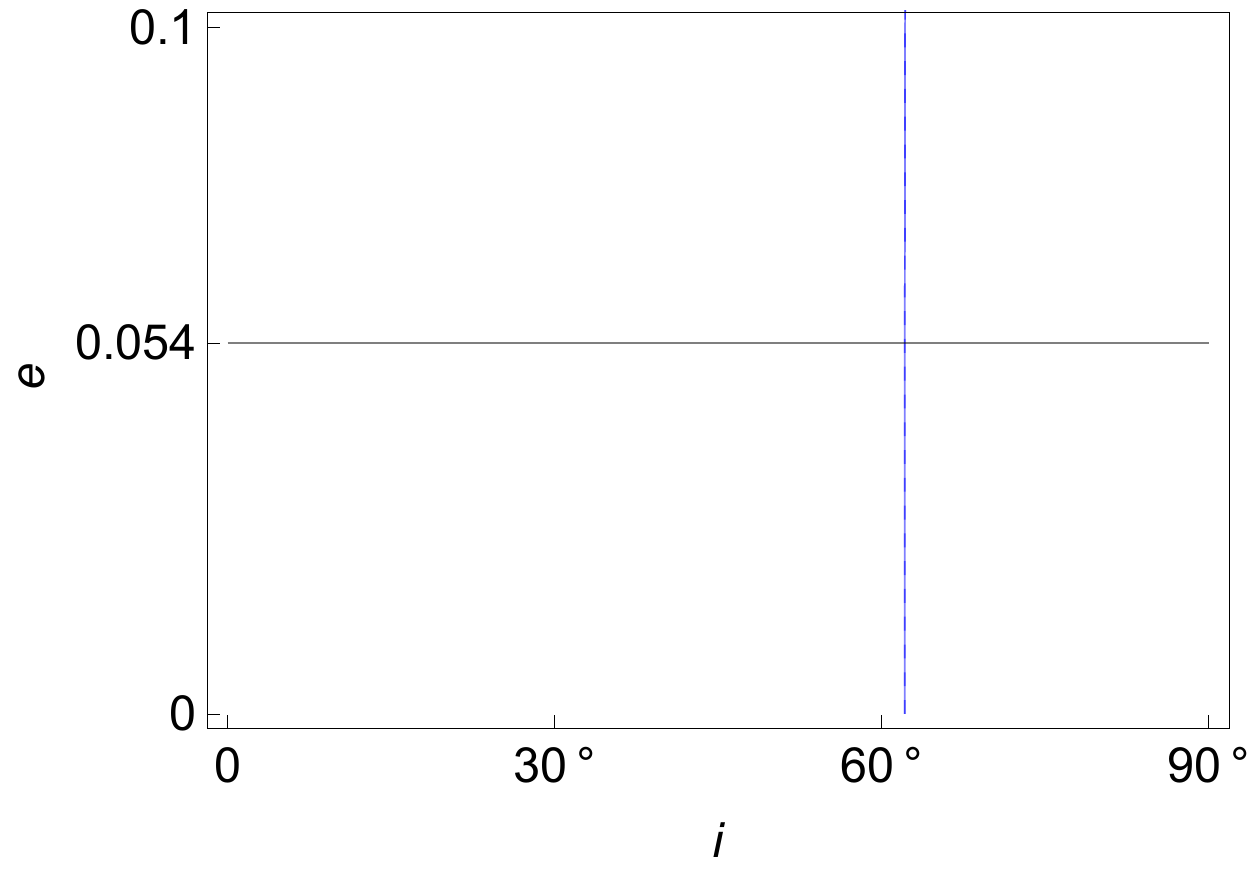}}
\subfigure[\label{figure:fig4b} $\lbrack J_2-J_3 \rbrack$.]{\includegraphics{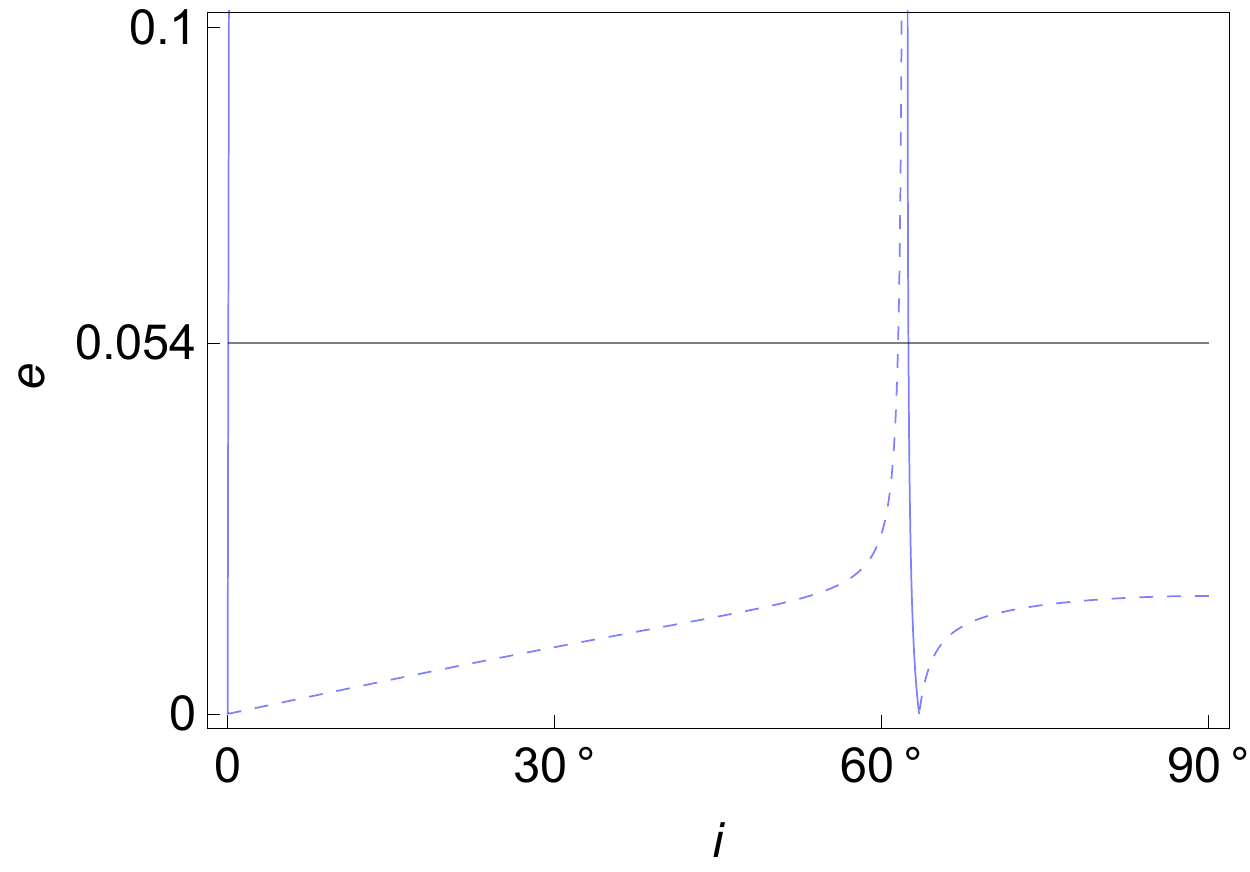}}
\subfigure[\label{figure:fig4c} From $\lbrack J_2-J_4 \rbrack$ to $\lbrack J_2-J_6 \rbrack$ in red; $\lbrack J_2-J_7 \rbrack$ in blue.]{\includegraphics{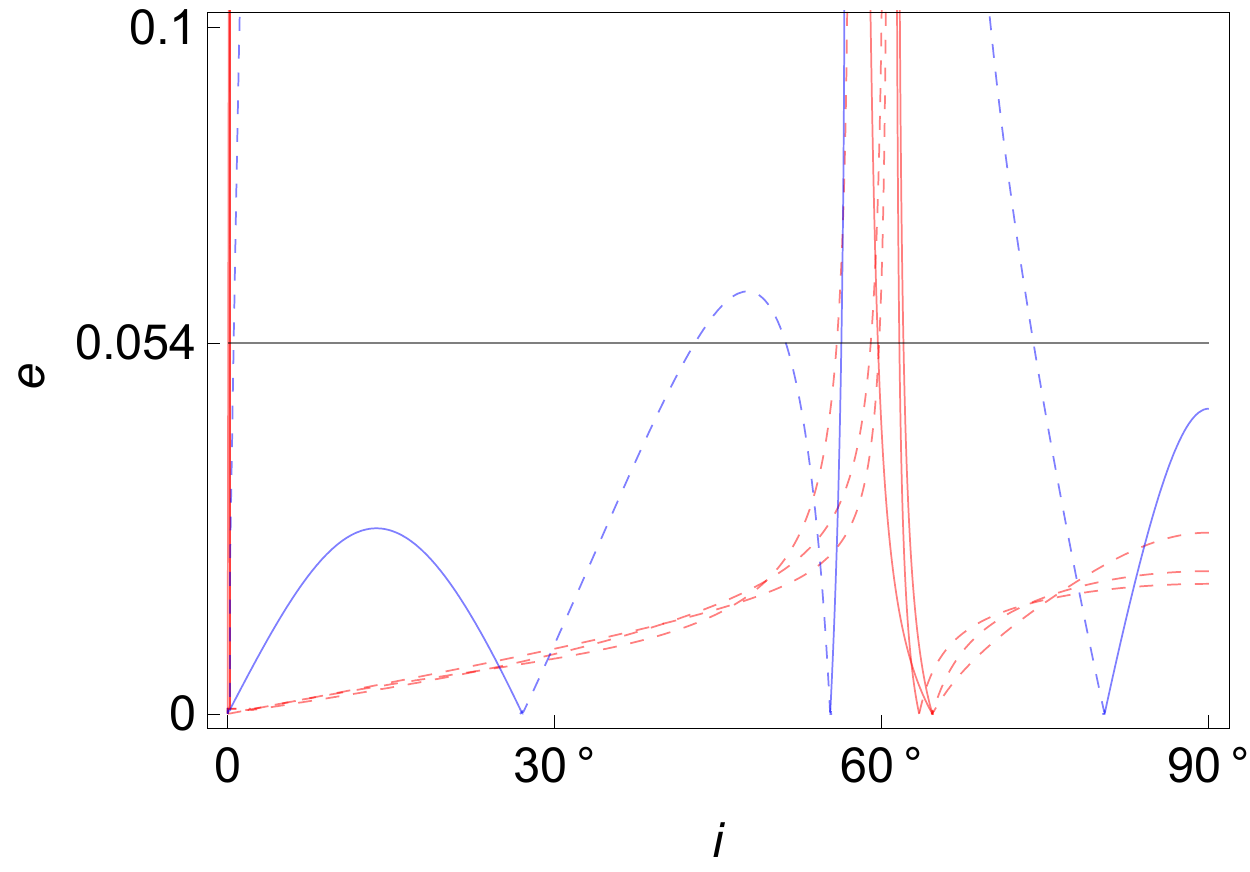}}
\subfigure[\label{figure:fig4d} $\lbrack J_2-J_8 \rbrack$ in red; $\lbrack J_2-J_9 \rbrack$ in blue.]{\includegraphics{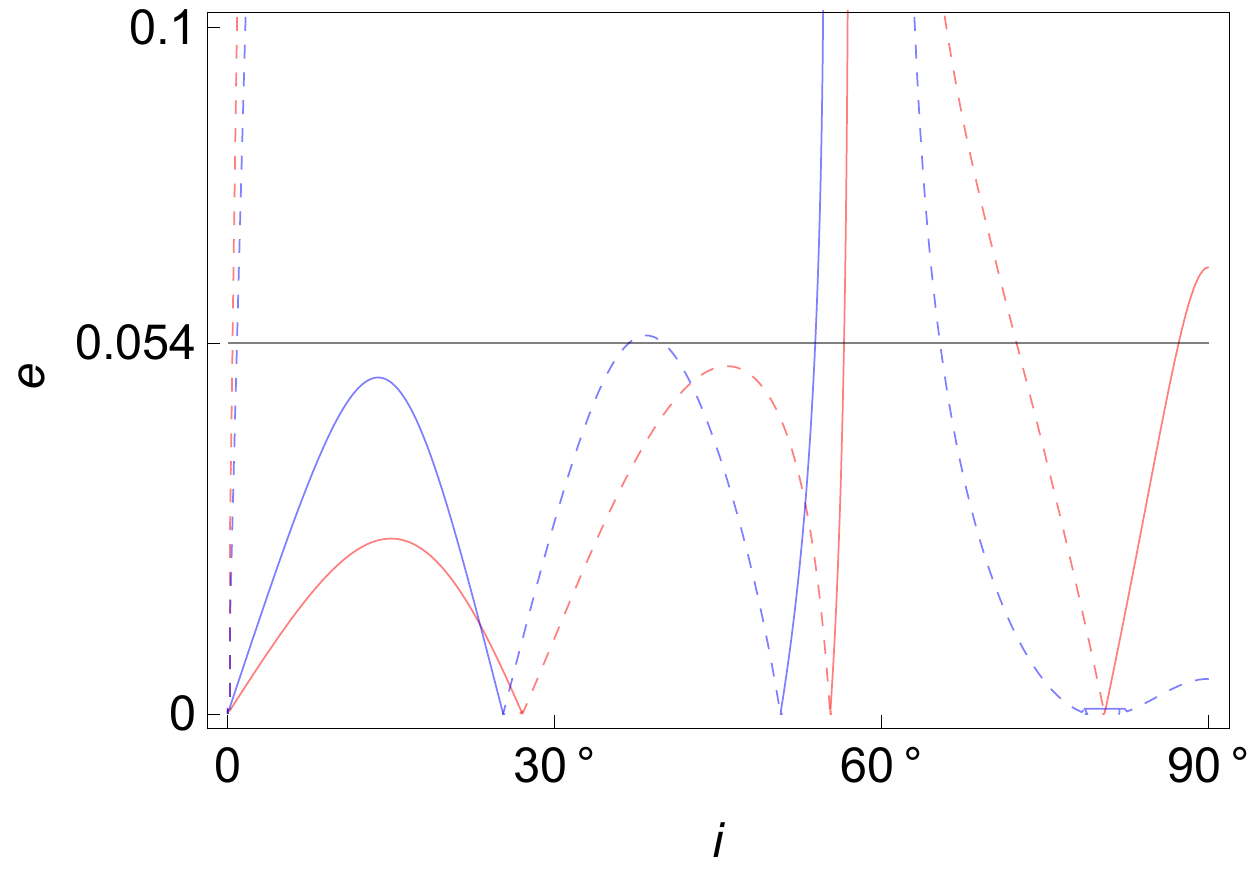}}
\end{subfigmatrix}
\caption{Variations in the number of families of frozen orbits for a mean semi-major axis of $\bm{1838}$ km.}
\label{figure:fig4}
 \end{figure}

Finally, with the aim of highlighting the importance of considering the high-order harmonics of the lunar gravitational field for the purpose of searching for frozen orbits in real applications, we delve into the effect that each of the zonal harmonics contributes to the characterization of frozen orbits. A similar approach has been recently proposed in \cite{lar2018gru_truncfroz_stardust} to ascertain the correct truncation of the Geopotential required in orbit prediction problems. We plot separately the families of frozen orbits that correspond to the joint effect of the lunar gravitational field up to each individual order from $2$ to $50$ and the third-body attraction from the Earth. In general, every new harmonic that is included modifies very slightly the distribution of frozen orbits from lower-order terms. Nonetheless, there are three specific harmonics, $J_3$, $J_7$, and $J_9$, that introduce variations in the number of families of frozen orbits. Then, in each of the graphics of Fig.~\ref{figure:fig4} we gather together the plots that correspond to zonal consecutive harmonics with similar behavior, and include in blue the plot with the next harmonic that introduces a change in the number of families of frozen orbits. The same conventions from previous figures regarding the argument of the periapsis and the eccentricity limit for non-impact orbits also apply to Fig.~\ref{figure:fig4}.

Figure~\ref{figure:fig4a} shows a single critical inclination when only $J_2$ is considered. If $J_3$ is also included, Fig.~\ref{figure:fig4b}, new frozen orbits appear with an argument of the periapsis of $270^\circ$. The incorporation of the $J_7$ zonal harmonic in Fig.~\ref{figure:fig4c} implies that a new circular frozen orbit appears near the inclination of $30^\circ$, whereas the one that already existed near $60^\circ$ is shifted towards a higher inclination, near $80^\circ$. As a consequence of the former, a new family of frozen orbits appears for an argument of the periapsis of $90^\circ$, whereas the latter implies that a wide region of inclinations between approximately $60^\circ$ and $75^\circ$ is deprived of feasible frozen orbits with eccentricities under the impact limit. In addition, the change in the argument of the periapsis that this circular frozen orbit caused before, from $90^\circ$ to $270^\circ$, now becomes the opposite, that is, from $270^\circ$ to $90^\circ$. Finally, Fig.~\ref{figure:fig4d} reveals a new family of frozen orbits surrounding the inclination of $80^\circ$ when $J_9$ is taken into account.

\begin{figure}[!!htp]
\begin{subfigmatrix}{2}
\subfigure[\label{figure:fig5a} From $\lbrack J_2-J_{10} \rbrack$ to $\lbrack J_2-J_{27} \rbrack$ in red; $\lbrack J_2-J_{28} \rbrack$ in blue.]{\includegraphics{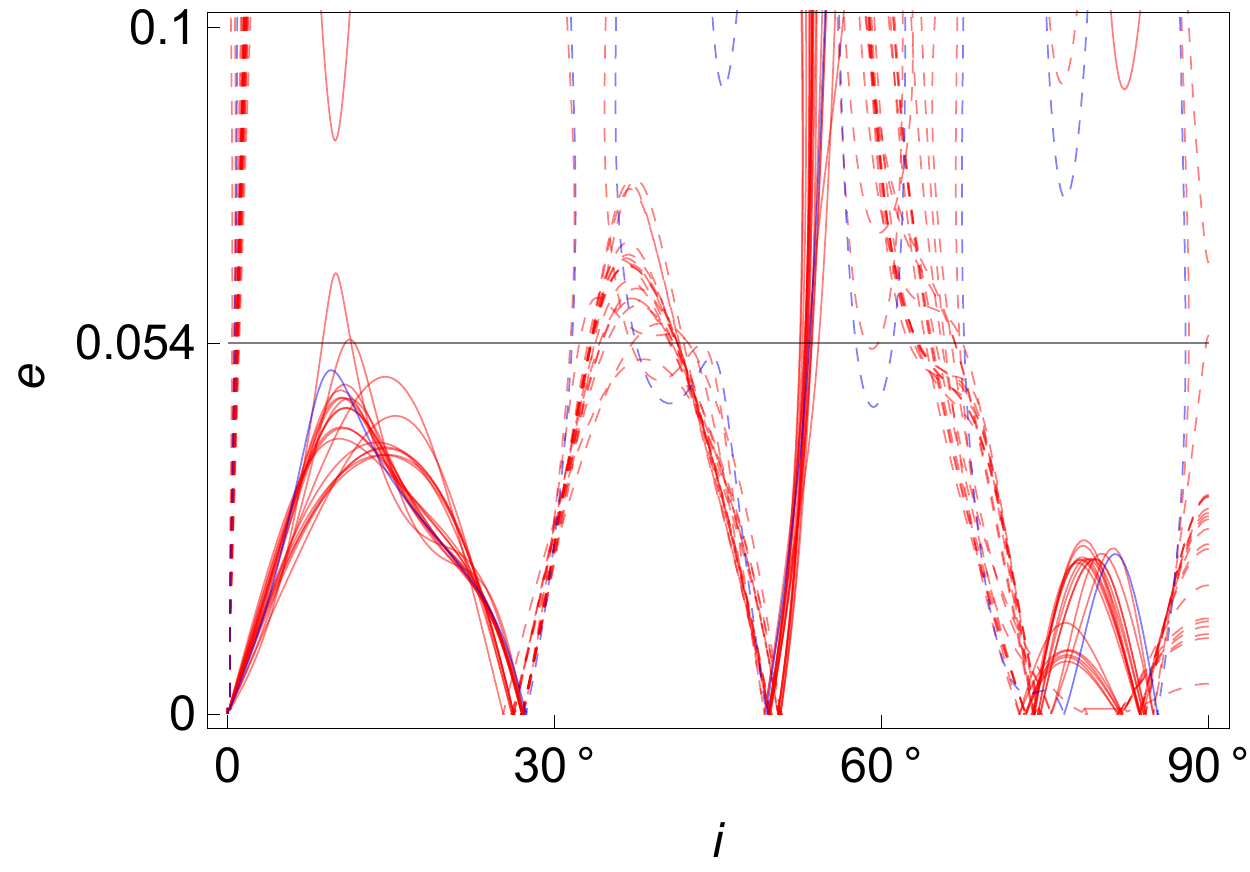}}
\subfigure[\label{figure:fig5b} From $\lbrack J_2-J_{29} \rbrack$ to $\lbrack J_2-J_{49} \rbrack$ in red; $\lbrack J_2-J_{50} \rbrack$ in blue.]{\includegraphics{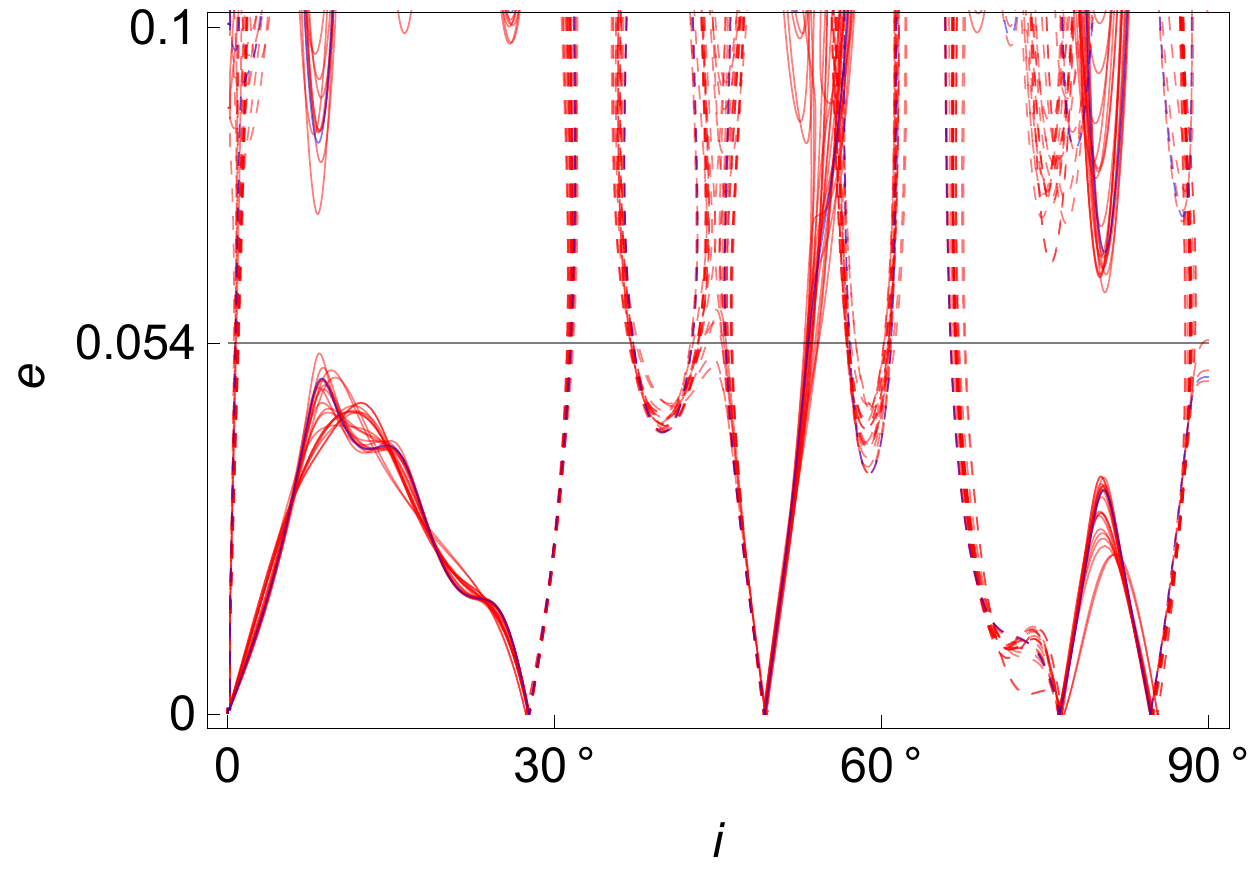}}
\end{subfigmatrix}
\caption{Evolution of the different families of frozen orbits for a mean semi-major axis of $\bm{1838}$ km.}
\label{figure:fig5}
 \end{figure}

Higher-order harmonics over $J_9$ do not modify the number of families of frozen orbits, although they introduce variations in their evolution. In order to illustrate those changes, we divide the plots that correspond to the incorporation of every new harmonic into Figs.~\ref{figure:fig5a} and \ref{figure:fig5b} for the sake of clarity, following the same conventions as in Fig.~\ref{figure:fig4}. The most important variations concern the feasibility of certain frozen orbits, which can cross the eccentricity limit for non-impact orbits in either direction when new harmonics are taken into account. That is especially significant for polar orbits, which, depending on the number of considered harmonics, can be frozen or not.

This analysis evinces the necessity of considering high-order lunar gravitational models when searching for low-altitude frozen orbits as part of a real mission-design process. Otherwise, certain families of frozen orbits will be missed, and the rest will be characterized with inaccurate values of eccentricity and inclination, especially in the region of polar orbits, to the extent that ignoring high-order harmonic coefficients can regard an impact orbit as a feasible frozen orbit. Therefore, it is essential to highlight that analyses that restrict their gravitational models to a few harmonics can only constitute academic studies, but can never be recommended for real applications.

\section{Conclusion}

We have conducted this study in order to clarify what the minimum resolution of the lunar gravitational potential should be considered when searching for frozen orbits in a real mission-design context, rather than within an academic scope. With this aim in view, we have developed a closed-form second-order semi-analytical theory for low-altitude orbiters around the Moon by applying perturbation methods based on Lie transforms. A $50\times 50$ lunar gravitational field has been considered, together with the third-body attraction from the Earth. The application of Lie transforms allows removing the short- and medium-period terms. Once an analytical theory is available, the full phase space of the dynamical system can be described, which allows, among other applications,  characterizing the different families of frozen orbits. That can constitute a valuable tool for searching for frozen orbits in mission analysis and design, provided a reliable force model is used. In this study we have restricted the search for frozen orbits to the well-known argument-of-the-periapsis values of $90^\circ$ and $270^\circ$. In principle, frozen orbits could also exist for other arguments of the periapsis, although, given the complexity of the expressions to manipulate, finding them can be a formidable task.

We have analyzed in depth the effect that changing the number of lunar harmonic coefficients that are considered produces on the distribution of families of frozen orbits. It has been verified that an insufficient order can miss certain families, and, what is worse, may predict the existence of valid frozen orbits at certain inclinations at which no frozen orbits exist for eccentricities below the impact limit. We have concluded that a $50\times 50$ gravitational model should be the minimum for real applications, thus confirming previous results in the scientific literature obtained with different techniques. Lower orders can only be regarded as academic studies, whereas in the case of higher orders, since the qualitative behavior no longer changes, the small gain in accuracy does not justify the significant increase in the computational load.

Likewise, the influence of the gravitational attraction from the Earth has also been analyzed. It can be concluded that the effect of this perturbation on the distribution of families of frozen orbits is especially important for orbits near the polar region, at inclinations above approximately $70^\circ$, where this third-body effect allows for the existence of viable frozen orbits that, otherwise, would not be possible for eccentricities below the impact limit.

Finally, we have implemented this semi-analytical theory in a \textit{Mathematica} package for mission planning, which allows finding suitable frozen conditions that constitute an accurate first approximation for a periodic orbit in the original space, once mean elements have been converted into osculating elements, that is, once the short- and medium-period terms have been recovered.

\section*{Funding Sources}

This work has been funded by the Spanish State Research Agency and the European Regional Development Fund under Project ESP2016-76585-R (AEI/ERDF, EU).

\section*{Acknowledgments}

The authors would like to thank two anonymous reviewers for their valuable suggestions.

\bibliography{references}

\end{document}